\documentstyle[preprint,aps,eqsecnum]{revtex}

\tighten
\def\be{\begin{equation}}
\def\ee{\end{equation}}
\def\bea{\begin{eqnarray}}
\def\eea{\end{eqnarray}}
\def\vecs{{\bf S}}

\begin{document}

\draft

\title{Reexamination of the microscopic couplings of the quasi 
one--dimensional antiferromagnet CuGeO$_3$}

\author{K. Fabricius, A. Kl\"umper, U. L\"ow}
\address{Fachbereich Physik, Bergische Universit\"at Wuppertal, 
Gau\ss str. 20, 42097 Wuppertal, Germany}
\address{Institut f\"ur Theoretische Physik, Universit\"at zu K\"oln,
Z\"ulpicher Str. 77, 50937 K\"oln, Germany}
\address{Institut f\"ur Physik, Universit\"at Frankfurt,
Robert-Mayer-Str. 2-4, 60325 Frankfurt, Germany}
\author{B. B\"uchner, T. Lorenz}
\address{II. Physikalisches Institut, Universit\"at zu K\"oln,
Z\"ulpicher Str. 77, 50937 K\"oln, Germany}
\date{\today}

\maketitle

\begin{abstract}
Experimental data for the magnetic susceptibility and magnetostriction 
of $CuGeO_3$ are analyzed within a one-dimensional 
antiferromagnetic model with nearest (${J_1}$)
and next-nearest neighbour interactions (${J_2}$).
We show that the ratio of the exchange constants 
in the antiferromagnetic chains
of $CuGeO_3$ amounts to ${J_2}/{J_1}$ = 0.354(0.01), i.e.
it is significantly larger than the critical value for the formation
of a spontaneous gap in the magnetic excitation spectrum without lattice
dimerization. The susceptibility data are reproduced by our numerical
results over the temperature range from 20K to 950K to a high degree 
of accuracy for ${J_1} = 80.2 (3.0)$ and ${J_2} = 28.4 (1.8)$. The 
pressure dependence of the exchange constants is estimated from 
magnetostriction data. Furthermore, the specific heat data are checked 
on consistency against the calculated entropy of the above model.
\end{abstract}


\section{Introduction}

The investigation of low dimensional quantum spin systems
has attracted widespread and general interest in active
research on a large class of magnetic materials,
both experimentally as well as theoretically, since the
properties are strongly affected by quantum fluctuations.
In particular, there is strong theoretical and experimental
effort to understand the origin of singlet-triplet spin
gaps in low dimensional spin systems such as $Sr Cu_2O_3$
and $VO_2P_2O_7$, which occur in the presence of competing
spin--interactions, e.g. due to a spin ladder geometry.

Evidence for frustrated spin--interactions
has also been reported for the quasi--one--dimensional
antiferromagnet $CuGeO_3$~\cite{Riera95,Castilla95,Buechner96}.
Initially this compound has received much interest,
since it is the first example of an inorganic compound undergoing
a spin--Peierls transition~\cite{Hase93}.
Subsequent extensive experimental studies
have revealed that many properties of the ordered phase
are well described by the well known spin--Peierls 
scenario~\cite{Bray76,Bulaevski78,CF,Pytte},
such as the presence of a lattice dimerization~\cite{Hirota94}
and a singlet triplet spin--gap scaling with
the lattice distortion~\cite{Nishi94}.
However, recently it was
found that the formation of the non--magnetic low temperature
phase in $CuGeO_3$ sensitively depends on details of the magnetic
exchange \cite{Buechner96}. 

The traditional 
spin--Peierls theory~\cite{CF,Bulaevski78}
is based on one--dimensional antiferromagnetic
chains with nearest-neighbour couplings only. Such magnetic systems
do not show any long-range antiferromagnetic order in the groundstate,
however they possess critical quantum fluctuations which drive the
lattice dimerization, i.e. the spin-Peierls transition. Substances with 
frustrated spin interactions are not strictly governed by 
\cite{CF,Bulaevski78} as the magnetic system in these cases 
may show spontaneous long-range magnetic dimerization in the groundstate
even without any lattice dimerization. The 
additional spin-phonon coupling merely stabilizes this magnetic dimerization
upon developing the lattice distortion. Whether this or the former scenario 
is realized depends on the strength of the frustration parameter 
$\alpha=J_2/J_1$. 

Frustration of the spin--interactions in $CuGeO_3$ has been
inferred previously from
the investigation of the magnetic susceptibility ($\chi$) in the
non-dimerized phase~\cite{Riera95,Castilla95},
which is in disagreement with a nearest neighbour Heisenberg 
model~\cite{Hase93,Castilla95,Riera95}.
A much better agreement has been found in theoretical studies of the
spin--susceptibility $\chi(T)$ invoking a
Heisenberg chain with competing nearest and next-nearest neighbour
exchange couplings ${J_1}$ and ${J_2}$ ~\cite{Castilla95,Riera95}.

However, two markedly different choices of exchange couplings, i.e.
$({J_1},\alpha) = (75K,0.24)$~\cite{Castilla95} and 
$({J_1},\alpha) = (80K,0.36)$~\cite{Riera95}, 
were derived within model calculations
and the {\em same} experimental data for the
magnetic susceptibility as well as inelastic
neutron scattering. For these quantities
an increase of $\alpha$ on one hand and
a decrease of ${J_1}$ on the other have similar
consequences leading apparently to a large uncertainty
of the exchange parameters.

Thus these previous studies of the quantum magnetism
in $CuGeO_3$ reveal some evidence for the
relevance of magnetic frustration in $CuGeO_3$, whereas
it is obviously difficult to extract precise values
of the exchange parameters.
A precise knowledge of the ratio $\alpha$ is of course very
important, since the theory predicts a critical
ratio $\alpha_c$ for a spin--gap to develop in the
magnetic excitation spectrum. This gap
opens irrespective of a lattice distortion.
The existence of this gap is established exactly at the Majumdar--Ghosh point
$\alpha=1/2$ \cite{Majumdar69} and by several numerical studies
\cite{j1j2theor,Castilla95,Riera95} which strongly
suggest $\alpha_c\approx 0.2411$.

In order to decide whether the spin gap in $CuGeO_3$
is at least partially a manifestation of
frustrated spin--interactions in a low dimensional
magnet we have performed a comparative study of theoretical results and
experimental data for thermodynamic properties.
Using our numerical results for the Heisenberg chain with nearest 
and next-nearest neighbour interaction we show that a
comparison of the magnetic susceptibility alone allows for the 
unique determination of the exchange constants in the frustrated
one-dimensional magnet $CuGeO_3$ with result
$\alpha = 0.354 \pm 0.01$, i.e. a frustration significantly larger
than the critical value $\alpha_c$.

In the course of our investigations we also determine the pressure dependence 
of the coupling parameters from magnetostriction data. We also analyse the
experimental specific heat data on consistency with the theoretical
entropy results.

\section{theory and numerics}

The dominant magnetic interactions in $CuGeO_3$ are due to Heisenberg 
spin exchange between $Cu^{2+}$ ions along the $c$-axis of the crystal. The 
Hamiltonian for the spin chain in the non-dimerized phase reads
\be
H=2\sum_i(J_1 \vecs_i \vecs_{i+1}+J_2 \vecs_i \vecs_{i+2}),
\ee
where we have adopted the normalization factor of \cite{BonnerF64}.
The nearest neighbour coupling $J_1$ is induced by the exchange 
path $Cu-O-Cu$, and the next-nearest neighbour coupling $J_2$ is caused by
the path $Cu-O-O-Cu$. A microscopic calculation of $J_{1,2}$ is difficult
\cite{Khomskii} and independent derivations are important.

In order to obtain quantitative results we calculate various 
physical properties in dependence on the 
couplings $J_1$ and $J_2$, 
notably the magnetic susceptibility, and perform a 
two-parameter fit of the experimental data which have been measured
up to 950K.
The aim is to achieve a best fit 
above the transition temperature $T_{SP}=14.3 K$ within the model
of a magnetic system with interactions described above and an adiabatic 
decoupling of the spin-phonon interactions. In this sense $J_1$ and
$J_2$ are treated as effective coefficients explicitly dependent on the 
lattice geometry, i.e. microscopic bond angles and lattice constants.
Unfortunately, analytic results for the thermodynamics of the 
model are available only for $\alpha=0$ (nearest-neighbour Heisenberg 
chain). We therefore resort to complete numerical diagonalizations
of finite systems with chain lengths up to $L=18$. In general the numerical 
treatment of strongly correlated quantum spin chains is plagued by finite-size
effects at \it low \rm temperatures. Here, however, we are interested in 
relatively high temperatures with $k_B T > 0.5 J_1$. A comparison of 
numerical data for successive chain lengths $L=16, 17, 18$ shows that finite
size corrections are essentially negligible for our purposes. In Figs.
\ref{susa} and \ref{speza}
numerical results for the magnetic susceptibility and specific heat per
lattice site are depicted. Note the characteristic dependence of the 
extremal values $\chi_{max}$ and the corresponding $T_{max}$ on the 
frustration parameter $\alpha$. $T_{max}$ is decreasing with increasing
$\alpha$, whereas $\chi_{max}$ is increasing. The behaviour of the specific
heat is similar, however its maximal value $C_{max}$ is a decreasing function
of $\alpha$. Consequently, at low temperatures the entropy strongly increases
with increasing $\alpha$.

We first observe that the experimental susceptibility data $\chi$ allow for an 
unambiguous  determination of $J_1$ \it and \rm $J_2$ if we use the 
position $T_{max}=56$ K and absolute value $\chi(T_{max})$ of the maximum 
of $\chi$. In practice, we determine for a sequence of frustration
parameters $\alpha=J_2/J_1$ the value of $J_1$ leading to $T_{max}=56$ K.
For this set of coupling parameters the value of $\chi_{max}$ is calculated,
see Fig.\ref{Fig3}. The experimental value for $\chi_{max}$ is obtained in this plot
for $\alpha=0.354$ within an error of $0.01$. Note that 
$\alpha=0.24$ as used in \cite{Castilla95} would yield a value of 
$\chi_{max}$ far too large in comparison with the actually measured value.
The nearest neighbour coupling corresponding to $\alpha=0.354$
is $J_1=80.2 K \pm 3.0$. In Fig.\ref{Fig4}
the susceptibility $\chi$ for the values $\alpha=$ 0.354 and 0.24 is compared
with the experimental data.
Note the strong deviation of the theoretical
curve for $\alpha=0.24$ from the measured one especially at $T_{max}$. 
In contrast to this, the entire temperature dependence of $\chi(T)$ is 
reproduced very well for $\alpha=0.354$. The effect of finite-size corrections
on the numerical results has been reduced by a scaling analysis based on
a transfer matrix approach \cite{tobepublished}. The theoretical results are
reliable down to temperatures of 35 K.
Note that a Land\'e
factor of $g=2.256$ has been used as derived from ESR \cite{Honda96}. 
(Furthermore we have assumed a cancellation of van Vleck 
paramagnetism and core diamagnetism which is the most reasonable
assumption about the magnetic background. If we subtract a background of
$5\cdot 10^{-5} emu/mole$ the result would read $\alpha=0.362\pm0.01$.)
For the quoted error bars of $J_1$, $\alpha$ a relative
error of $3\%$ in the measurements has been taken into account. Also note
that a comparable analysis of the data \cite{Hase93} leads to 
$\alpha=0.371 \pm0.01$ ($\alpha=0.38 \pm0.01$ if a background of
$5\cdot 10^{-5} emu/mole$ is subtracted), 
a value which overlaps with the above result. The 
differences between our and the data in ref. \cite{Hase93} might originate
from slightly different orientations of the magnetic fields with respect
to the crystal axes. Note that our data agree much better with those in
ref. \cite{Pouget94}.

In addition to the strength of the microscopic couplings we can 
determine their pressure dependence. We obtain this from magnetostriction 
data \cite{Buechner96} which are related to the pressure dependence 
of the magnetic susceptibility \cite{Corr,Buechner96,Buechner97} 
which is rather directly accessible 
in numerical studies
\be
{1\over L_i}\left({\partial L_i\over\partial H}\right)_{p_i}=
{H\over V}\left({\partial\chi\over\partial p_i}\right)_{H}=
{H\over V}\left(\partial_1\chi
\left({\partial {J_1}\over \partial p_i}\right)_{T}
+\partial_2\chi
\left({\partial \alpha\over \partial p_i}\right)_{T}\right),
\ee
where $\partial_{1,2}$ denotes the derivative of $\chi$ with respect to $J_1$
and $\alpha$.
Based on this relation and the data for temperatures 40 K and 60 K we 
find ${\partial J_1/\partial p_i}=3.3(3),-7.5(5),-1.4(2)$
$\hbox{K GPa}^{-1}$, and
${\partial \alpha/\partial p_i}=-0.03(3),0.01(4),0.04(2) $
$\hbox{GPa}^{-1}$ for the 
three lattice axes $i=a$, $b$, and
$c$. Note that the values for the pressure dependence of $\alpha$ obtained 
along this way are consistent with the hydrostatic pressure dependence 
obtained in \cite{Loosdrecht}.
However, in our analysis the hydrostatic pressure dependence of $J_1$
is much stronger than that of $J_2$ which is essentially zero. 

\section{Entropy analysis}

Unfortunately it is impossible to measure the temperature dependence
of the magnetic specific heat in $\rm CuGeO_3$ directly. As displayed in
the left part of Fig.\ref{enta} at temperatures
close to the predicted maximum of $C_{mag}$
the total specific heat is dominated by the phonon contribution
$C_{ph}$, i.e. it is about one order of magnitude larger than
the calculated magnetic contribution. Therefore it is impossible
to extract $C_{mag}$ with sufficient accuracy to resolve the
small differences predicted for different exchange constants
in a $J_1-J_2$ model, see Fig.\ref{speza}. 
Note that this would require a knowledge of the phonon
contribution with an unrealistic accuracy of the order of
$10^{-3}$ or higher.
Nevertheless, one can use the measurements of $C$ to check different
exchange parameters suggested in the literature 
\cite{Hase93,Nishi94,Goetz,Kuroe,Gros97} for consistency.
In the following we will demonstrate that indeed most of these values for
$J_1$, $J_2$ lead to discrepancies.

It is well known that a reliable separation of magnetic and phonon
contributions of $C$ is possible at low temperatures 
\cite{Liu95,Weiden95,Ammerahl97} and we will
use this separation to estimate the magnetic entropy at higher
temperatures. Note that the spin--Peierls phase transition does not
modify $C_{ph}$ significantly, i.e. the anomaly of $C$ is due
to the magnetic contribution.
This is inferred on one hand from the extremely small structural
changes at the phase transition
and on the other hand consistent with the findings in measurements 
as functions of magnetic fields and 
doping~\cite{BBTLunpublished,Ammerahl97}.

Analyses of the specific heat at low temperatures
have been reported several times \cite{Liu95,Weiden95,Ammerahl97}. 
Well below $T_{SP}$ $C_{mag}$ shows
activated behavior due to the large spin gap.
Moreover, at low temperatures, i.e.
at and below $T_{SP}$, the specific heat and correspondingly the
entropy are dominated by their magnetic contributions.
At low temperatures the phonon
contribution follows the usual $T^3$ law, i.e. $C_{ph} = \beta T^3$
with $\beta \simeq 0.3mJ/K^4mole$.
At higher temperatures $C_{ph}$ deviates from this
$T^3$ behavior as illustrated in Fig.\ref{enta}. The extrapolation
of the low temperature behavior exceeds the total specific heat
already at moderate temperatures of about 30K. Therefore it may
serve as an upper limit for $C_{ph}$, which we
use to derive a lower limit $C_{mag}^{min}$ for the magnetic
contribution from the data.

Below 20K we determine this lower limit simply from the
difference $C-\beta T^3$ by assuming a large value 
of $\beta \simeq 0.32mJ/K^4mole$.
The difference function shows a maximum well above $T_{SP}$ at 20K. 
A further extrapolation
of this function to higher temperatures leads to
a decrease which is unreasonable for an estimate of $C_{mag}$. We therefore
take the value of $C-\beta T^3$ at 20K as 
a lower limit of $C_{mag}$ at temperatures above 20K (see Fig.\ref{enta}).
Of course this treatment yields only a very small
lower limit for $C_{mag}$, which is indeed much smaller
than the predicted $C_{mag}$ (see Fig.\ref{enta}a).
We do not aspire a more realistic extraction of $C_{mag}$
here, since on one hand it is impossible
to achieve the necessary accuracy to fix the exchange constants
and on the other hand the very conservative estimate
of $C_{mag}^{min}$ already suffices to rule out certain 
exchange constants reported for $\rm CuGeO_3$.

>From the lower limit of $C_{mag}$ we have calculated the
corresponding minimum magnetic entropy
$S_{mag}^{min}$, which is compared to calculations of
$S_{mag}$ for several choices of $J_1$ and $J_2$ in the right part of
Fig.\ref{enta}. Taking the temperature at the maximum of $\chi$ and assuming
$\alpha=0$ yields $J=44K$
as described in the initial paper on
the spin--Peierls transition in $CuGeO_3$ \cite{Hase93}. 
The corresponding entropy is much larger than the lower limit
we have extracted from the data. Nevertheless, the specific heat
data allow to exclude these parameters. At temperatures slightly above
$T_{SP}$ the calculated magnetic entropy for $J_1=44K$, $J_2=0$
amounts to about 95\% of the total entropy in
$CuGeO_3$. In other words
an unreasonably small phonon background, e.g. with a $\beta \simeq 0.04
mJ/moleK^4$ which is nearly one order of magnitude too small,
has to be chosen to account for this large magnetic entropy.

A further suggested value for the intrachain exchange constant
in a model with $\alpha=0$ is
$J = 60K$. This value has been reported first in \cite{Nishi94}
based on their inelastic neutron scattering data. More
recently in \cite{Goetz} the same value and in addition
a large (frustrated) intrachain exchange was found from
the analysis of the dispersion curves in the dimerized phase, i.e. below
$T_{SP}$.
It is apparent from Fig.\ref{enta} that the magnetic entropy calculated
for these exchange parameters in a one-dimensional model,
i.e. for $J_1 = 60K$, $J_2=0$ and above $T_{SP}$, is significantly
smaller than the lower limit $S_{mag}^{min}$ extracted from the data.
Thus the specific heat does not support
the parameters suggested in ref.~\cite{Goetz}, in particular when
taking into account a further reduction of the theoretical $S_{mag}$
due to the suggested large interchain exchange. 

The three solid lines in the right part of Fig.\ref{enta} correspond to
calculations of $S_{mag}$ for different exchange constants
and finite $\alpha$. Since these entropies have been calculated
for finite chains, the results are reliable for
temperatures above approx. 35K.

The strongest discrepancy between numerical
and experimental data is present for
the set of parameters $J_1 = 125K, \alpha = 0.35$.
These parameters have been suggested recently 
in \cite{Kuroe} to give the best description of the
magnetic specific heat in $CuGeO_3$. In this latter work
$C_{mag}$ has been extracted from Raman scattering data
and it was concluded that it is impossible to fit both
the susceptibility and the magnetic specific heat with
a single choice of $J_1$ and $J_2$.
However, it is apparent from Fig.\ref{enta} that the parameters
suggested in ref.~\cite{Kuroe} and consequently the ``magnetic specific
heat'' extracted from Raman scattering is in striking discrepancy
to the measurement of the specific heat at low temperatures. Thus our data
do not support the reported inconsistency 
in the determination of exchange parameters from $\chi$ and $C_{mag}$
\cite{Kuroe}.
Further investigations seem necessary to explain
the striking discrepancy between
$C_{mag}$ as revealed in \cite{Kuroe} from quasi--elastic scattering
and the true magnetic specific heat.

The deviation between the data and the calculations for the exchange constants
$J_1=75K, \alpha = 0.24$, which have been extracted by Castilla et
al. from their analysis of the magnetic susceptibility and the
dispersion curves, is less pronounced. 
The calculation of the magnetic entropy is reliable for temperatures
larger than 35K where the numerical results are indeed larger than
the lower bound we have estimated from the data.
A very nice convergence of theory and experiment appears
upon further increasing $\alpha$ to 0.354 (and J=80.2K). As shown
in Fig.\ref{enta} the
magnetic entropy calculated for the parameters which yield the best
fit to the susceptibility is always
larger than the lower bound extracted from the data.
Moreover, it is apparent from Fig.\ref{enta} that the difference
between the theoretical $S_{mag}(J_1=80.2K,\alpha={0.354})$ and the
lower bound systematically decreases with decreasing temperature, i.e.
with increasing accuracy of the extracted minimum magnetic
entropy. 

Unfortunately, it is impossible to extend
the theoretical calculations to temperatures of 20K and below.
At these temperatures the data for $S_{mag}$ shown in Fig.\ref{enta} do not only
represent a rough lower bound.
Close to $T_{SP}$
$S_{mag}$
is markedly larger than $S_{ph}$ for any
reasonable $C_{ph}$
(As mentioned above
the curve for $S_{mag}(J=44K)$ corresponds
already to 95\% of the total entropy). Therefore even
assuming a significantly
smaller phonon background leads only to moderate
changes of $S_{mag}$ at these temperatures.

Boldly extrapolating the calculated entropy for $J=80.2K,\alpha=0.354$
to lower temperatures one may conclude that
both the value of
$S_{mag}$ at $T_{SP}$ as well as its temperature dependence, i.e.
the magnetic specific heat, are well described.
In contrast to that for all other choices of
exchange parameters in Fig.\ref{enta} significant discrepancies between the
model calculations and the experimental data are apparent.
Though the accuracy for the determination of $C_{mag}$
is not sufficient to unambiguously determine the exchange constants
from these data alone,
the specific heat strongly confirms
our analysis of the susceptibility.
In particular, in contrast to the conclusions
of \cite{Kuroe} there is no evidence
that it is necessary to invoke
markedly  different
exchange constants
to explain $\chi$
and $C_{mag}$ (at low temperatures).

\section{Conclusion}

We have presented numerical results for thermodynamical properties of
a quantum spin-1/2 chain with nearest and next-nearest neighbour interactions
which is believed to be at the heart of the magnetic system of $CuGeO_3$.
The microscopic interaction parameters have been determined as well as their
uniaxial pressure dependence. We have 
shown that the frustration parameter is $\alpha=0.354$. This is much 
larger than the value used for the explanation
of Raman scattering data \cite{Gros97,Loosdrecht}. We expect
our result to be reliable as we have based our reasoning on established 
thermodynamical relations. Furthermore, we have demonstrated that the
experimental magnetic susceptibility data are accounted for in even 
quantitative details by the quasi one-dimensional model and $\alpha=0.354$. 

Within the present accuracy the magnetic specific heat calculated
for the exchange constants derived from our analysis 
of $\chi$ is consistent with the analysis of the
experimental data. On the other hand for several
other choices of exchange parameters which have been suggested
for $CuGeO_3$ we find not only a worse description
of the susceptibility but simultaneously evidence for
discrepancies to the specific heat data.

\begin{figure}
\caption[susa]{Plot of numerical results for the 
susceptibility per site (in units of $J_1$) versus the reduced 
temperature $T/J_1$ for $\alpha=$ 0.1, ..., 0.4, and analytical results
for $\alpha=0$ following \cite{Kl93} down to zero temperature.
Shown in insets: behaviour of the susceptibility 
in the neighbourhood of the maximal values $\chi_{max}$ and plot
of the corresponding 
temperature $T_{max}$ as a function of $\alpha$.}
\label{susa}
\end{figure}
\begin{figure}
\caption[speza]{Plot of numerical and analytical results for the 
specific heat per site  similar to Fig.1.}
\label{speza}
\end{figure}
\begin{figure}
\caption[]{Plot of the maximal value of the magnetic susceptibility $\chi_{max}$
as a function of $\alpha=J_2/J_1$ (solid line) in the two parameter model. 
The experimental value (dashed line) is crossed at
$\alpha=0.354$. (The corresponding value of $T_{max}$ is 56K.)}
\label{Fig3}
\end{figure}
\begin{figure}
\caption[]{Depiction of the experimental results for the
magnetic susceptibility in dependence on
temperature (solid line). Also shown are the theoretical results for 
$\alpha=0.354$ (dotted line) and $\alpha=0.24$ (dashed line).}
\label{Fig4}
\end{figure}
\begin{figure}
\caption[enta]{Left upper panel: Experimentally observed
specific heat of $\rm CuGeO_3$ ($\circ$). The
extrapolated low temperature phonon background $C_{ph} = \beta T^3$ with
$\beta = 0.32mJ/K^4mole$ is indicated by the dashed line. The solid
line shows the calculated magnetic specific heat for the exchange
constants revealing the best fit to the magnetic susceptibility.
Left lower panel: Experimentally observed
specific heat of $\rm CuGeO_3$ ($\circ$) and the estimated minimum
magnetic specific heat (solid line, see text).\\
Right panel: Comparison between the minimum
magnetic entropy as revealed from the specific heat data
($\bullet$) and calculations of $S_{mag}$ assuming
different exchange constants given in the figure.
The dashed lines correspond to results of exact thermodynamic calculations
for $\alpha=0$ and the solid lines are obtained from
numerical diagonalizations.}
\label{enta}
\end{figure}

\begin{references}
\bibitem{Corr} We mention that the uniaxial pressure dependences 
${\partial\chi\over\partial p_i}$ shown in Fig.2 of \cite{Buechner96}
and Fig.10 of \cite{Buechner97} are given in wrong units($10^{-7} emu/g GPa$).
The correct unit is $10^{-6} emu/cm^3 GPa$.
\end{references}
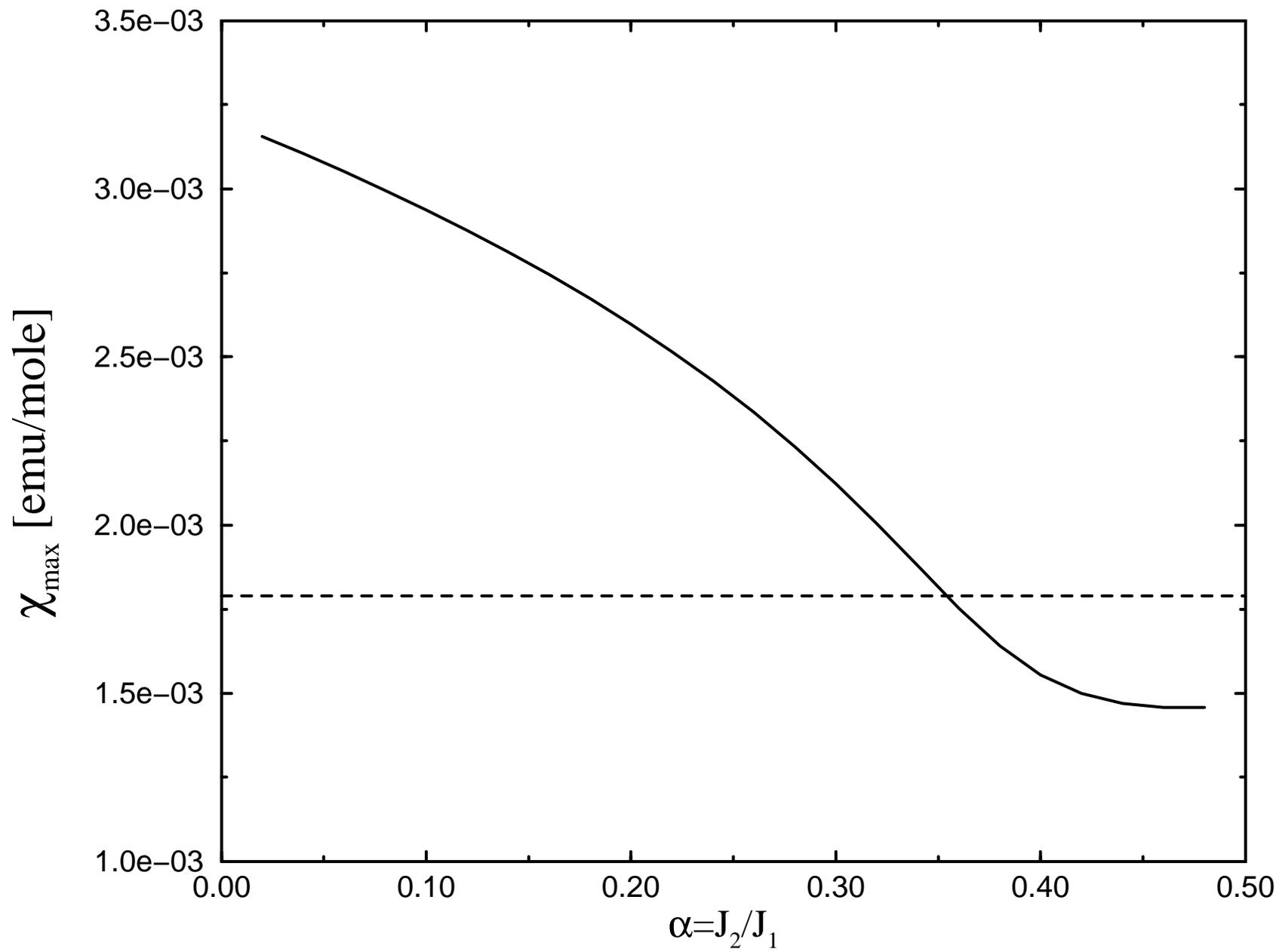
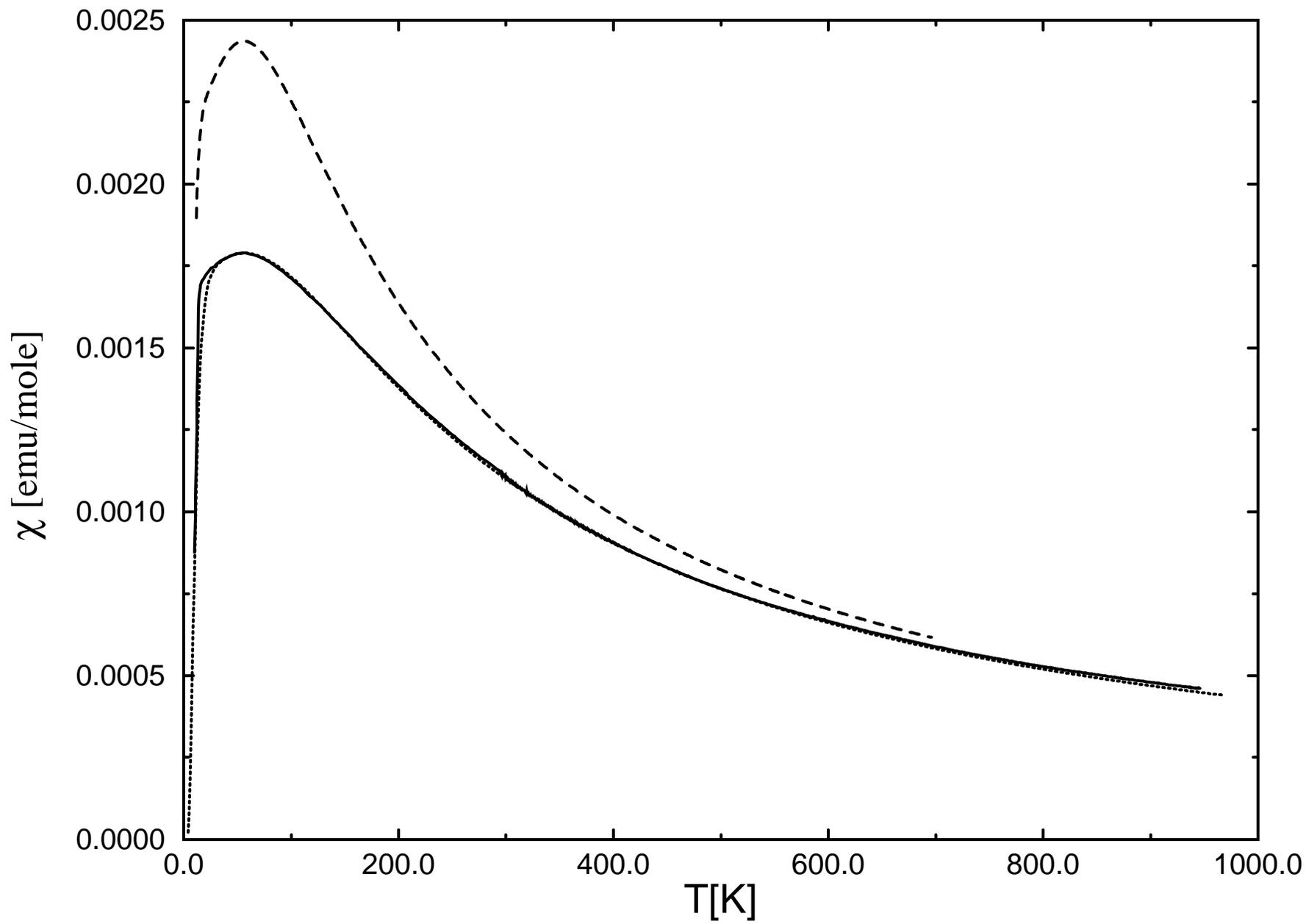
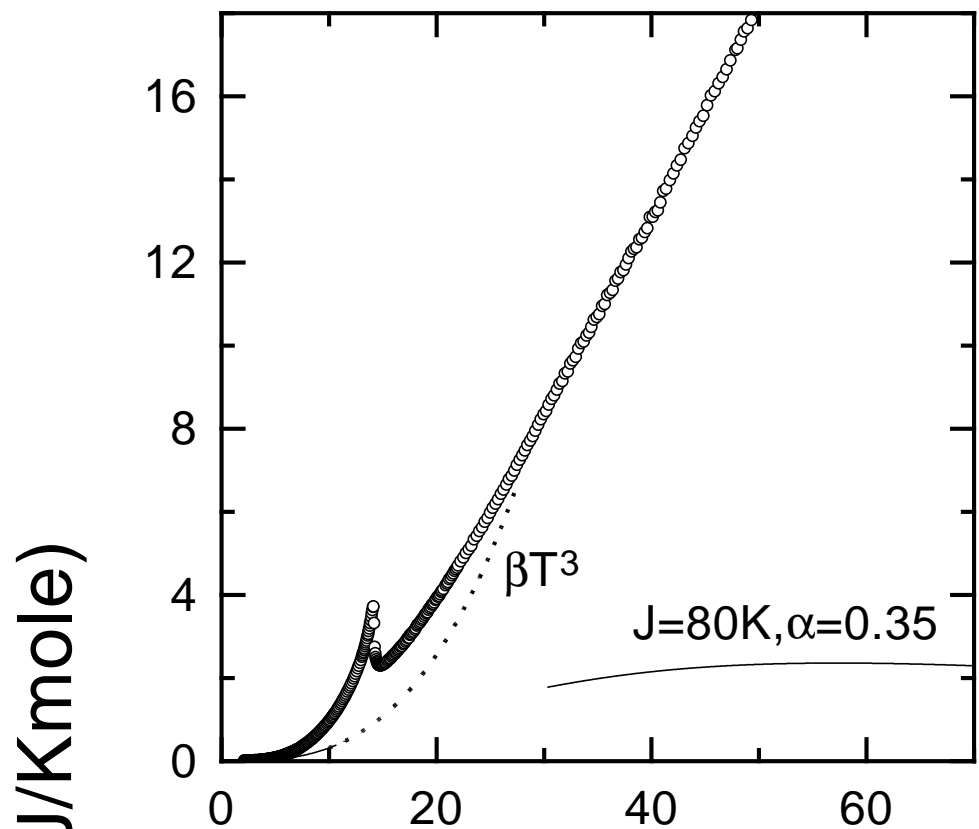
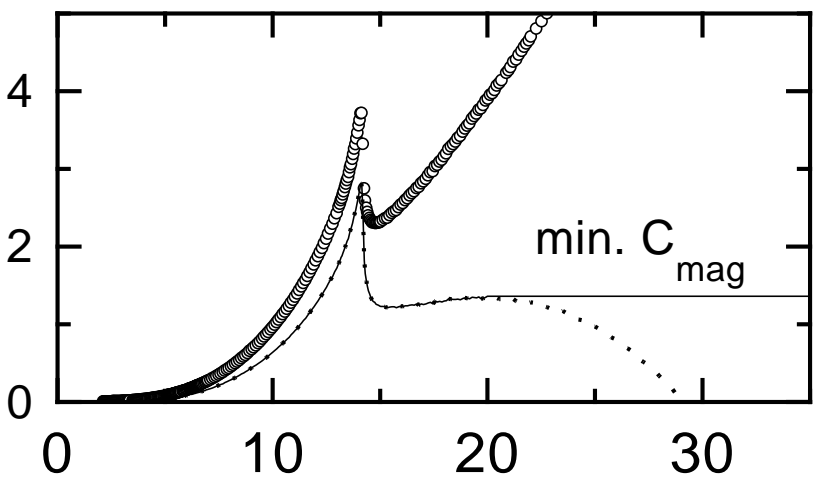
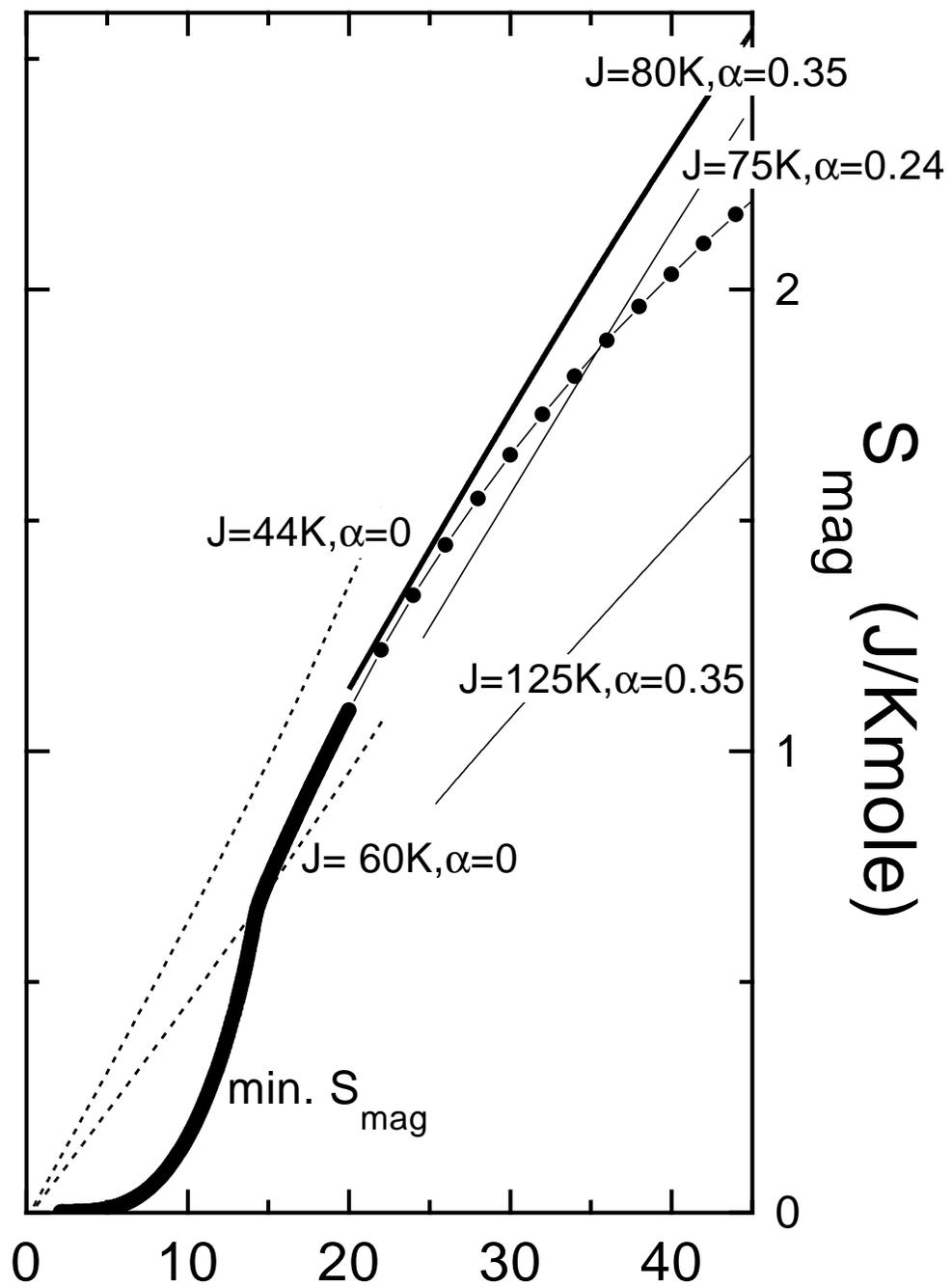
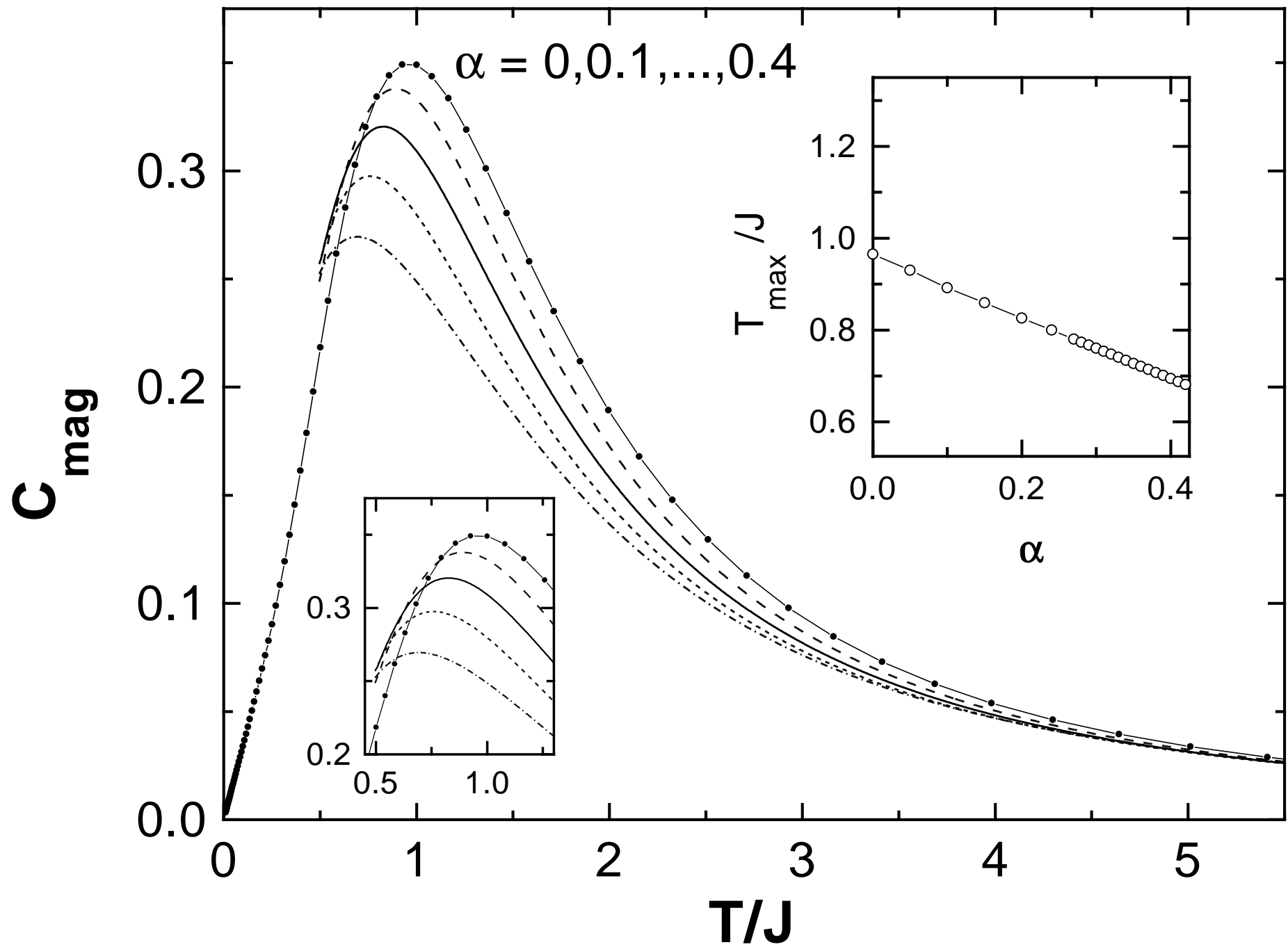
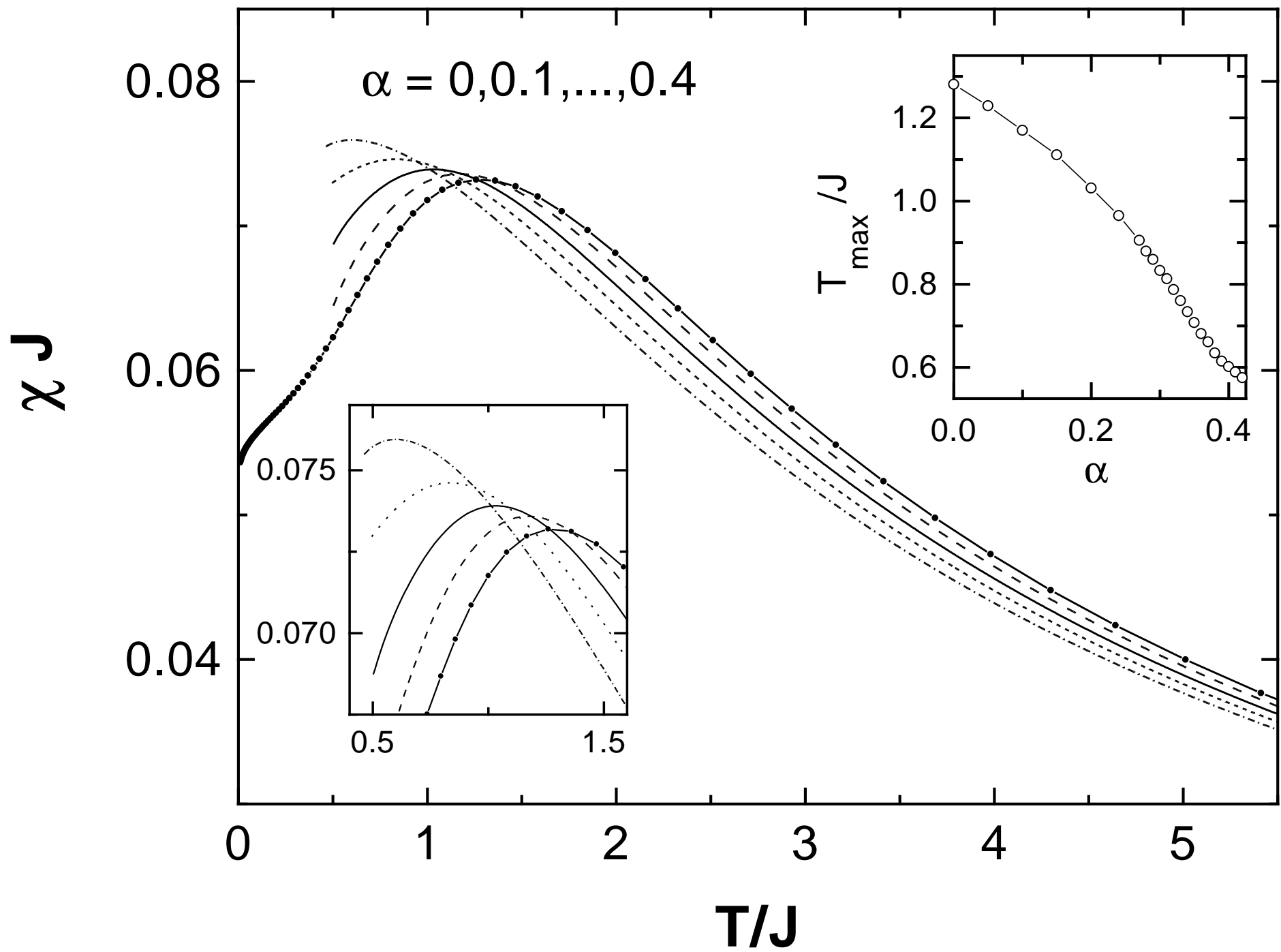
\end{document}